\documentstyle[aps,prl]{revtex}

\input {epsf}

\begin{document}
\draft
\preprint{draft}
\twocolumn[\hsize\textwidth\columnwidth\hsize\csname@twocolumnfalse\endcsname

\title{Hysteretic creep of elastic manifolds}

\author{Stefan Scheidl$^{(a)}$ and Valerii Vinokur$^{(b)}$}

\address{(a) Institut f\"ur Theoretische Physik, Universit\"at zu
  K\"oln, Z\"ulpicher Str. 77, D-50937 K\"oln, Germany\\ (b) Material
  Science Division, Argonne National Laboratory, Argonne, IL 60439}

\date{2 August 1996}

\maketitle

\begin{abstract}
  We study the dynamic response of driven systems in the presence of
  quenched disorder. A simple heuristic model for hysteretic creep of
  elastic manifolds is proposed and evaluated numerically. It provides
  a qualitative explanation of the phenomenology observed in
  experiments on high-temperature superconductors.
\end{abstract}


\vskip1pc]
\narrowtext

An intense attention is attracted currently by the driven dynamics of
vortices in high temperature superconductors.  This interest is
motivated not only by the technological request, but also because
vortex dynamics exhibits features which are generic for a diversity of
physical systems including charge density waves, Wigner crystals,
dislocation motion, polymers, and interfaces driven in quenched
disordered media.  In addition vortices are a unique experimental
object with tunable parameters. One of the fundamental features of
driven dynamics in quenched disorder is the existence of the so-called
{\it depinning transition}: at zero temperature the system cannot move
below a certain critical force whereas it moves above. At finite
temperatures the transition is smeared out but the depinning threshold
still marks the crossover from strongly non-linear thermally activated
dynamics dominated by pinning below the critical force to the viscous
flow above \cite{blatt}.

Pinning of the vortex lattice is governed by an interplay between
thermodynamic and kinetic effects. One of the manifestations of such
an interplay, the peak effect in the critical current, has been
recently observed in YBCO crystals prior to vortex lattice melting
\cite{kwokpeak}. Further investigations of the transport just below
the peak effect temperature range revealed another striking feature of
the depinning transition in the vortex lattice: dynamic hysteresis in
the I-V characteristic and aging effects \cite{exp}.  It was found
that the disappearance of the vortex motion (pinning) upon decreasing
the driving current and the depinning upon keeping the vortex system
at rest and subsequent current ramping occurs at different threshold
currents: depinning threshold appeared to exceed noticeably the
pinning current of the descending I-V curve.  Moreover, at an
ascending depinning threshold $j_{\rm th}$ the voltage jumps abruptly
from zero to the finite value $V(j_{\rm th})$ of the {\it descending}
branch of the I-V curve showing thus a strongly asymmetric hysteretic
behavior.  The value of the ascending threshold current was found to
depend on the waiting time $t_{\rm w}$ the system has spent at rest:
$j_{\rm th}$ grows with the $t_{\rm w}$ increasing.

The observed abrupt switching of the vortex motion resembles the
switching hysteretic behavior in charge density waves transport
\cite{cdw} and can be attributed to the formation of the bi-stable
dynamic state in the vortex motion, i. e. the dynamic state where both
weakly pinned and strongly pinned vortex motions can coexist.  The
formation of the bi-stable state is usually related either to the
possibility of plastic deformations (phase slips) of the
inhomogeneously pinned soft vortex lattice and/or to the formation
multivalued equilibrium state for pinned vortices. It has been
recently demonstrated that plastic deformations can become a
determining component of the depinning of the vortex lattice
\cite{plastic}. However the existing descriptions of plastic
deformation are lacking in aging and memory effects which are most
essential and characteristic for glassy dynamics. In this Letter we
will concentrate on this glassy aspect of the depinning process and
discuss plastic models elsewhere.

We consider a purely elastic threshold depinning process modeling the
vortex lattice as an elastic manifold and propose a phenomenological
equation describing a temporal evolution of the typical barriers
controlling the creep motion.  We demonstrate the existence of aging
effects which give rise to the abrupt increase in the manifold
velocity {\it within} the creep regime and to a dynamic hysteresis in
the depinning process. We find that the {\it threshold} force grows
with the increasing waiting time $t_{\rm w}$ the system spends at zero
force.  Our results provide a solid basis for a description of the
experimentally observed time-dependent hysteresis of the vortex
lattice threshold dynamic behavior \cite{exp} in terms of the aging
effect and predict the possibility of similar effects in all the
diversity of dynamic systems that can be modeled as an elastic
manifold in a random quenched environment.

We consider an overdamped dynamics of the elastic manifold subject to
the quenched disorder.  The character of motion follows from the
interplay between the (i) elastic forces that tend to keep the
manifold flat, (ii) pinning forces arising from the disorder potential
which pull the manifold into potential minima forcing thus {\it
  roughening} of the manifold, (iii) thermal forces of a heat bath at
temperature $T$, and (iv) a driving force $F$ which is homogeneous in
space but which may vary in time.

The structure of a manifold in equilibrium is determined by the
balance of elastic and pinning energies and can be characterized by
the difference correlation ({\it roughness})
\begin{equation}
  C(L)=\overline{\langle [u(L)-u(0)]^2 \rangle} \approx \xi_0^2
  (L/L_0)^{2 \zeta}
\label{C_0}
\end{equation}
of deformations $u$ from a flat configuration. $\xi_0$ denotes the
transverse correlation length of the disorder and $L_0$ is the Larkin
length. Disorder leads to a rough manifold with an exponent
$\zeta \geq 0$ below four dimensions \cite{rough}. The manifold
behaves as in a landscape of energy barriers of all sizes $L$ with
heights $U(L)$ which scale like\cite{hh,scal}
\begin{equation}
  U(L)= U_0 (L/L_0)^\theta.
\label{U(L)}
\end{equation}
The exponent $\theta=2(\zeta-\zeta_0)$ is related to the increase of
the roughness exponent $\zeta$ due to the presence of disorder
compared to the thermal roughness exponent $\zeta_0=(2-d)/2$. The
height of the smallest existing barriers on the Larkin scale is
denoted by $U_0$. 

The typical time to overcome a barrier of height $U$ by thermal
activation is
\begin{equation}
  \tau(U)= t_0 \exp[U/T]
\label{t(U)}
\end{equation}
with a microscopic temperature dependent time scale $t_0$. The
presence of barriers with unbounded heights gives rise to a glassy
behavior of the manifold with diverging relaxation times.  The central
point of our analysis is a self-consistent treatment of the the
dynamics of the manifold: we question the rate of the change of the
manifold position and configuration given by its structure $C(L)$,
taking into account that the barriers governing the dynamics are
determined themselves by the instant manifold configuration.

{\em Relaxation into equilibrium}.  We start with the relaxation of
the perfectly flat manifold, $C(L)=0$, into the equilibrium in the
absence of a driving force.  After a time $t$ the typical depth of the
metastable state the manifold falls in (or, equivalently, the height
of the barrier separating the typical metastable states) is
\begin{equation}
  U(t)=T\ln(t/t_0)
\label{U(t).relax}
\end{equation}
Therefore the manifold is equilibrated up to length scales 
\begin{equation}
  L_c(U) = L_0 (U/U_0)^{1/\theta}
\label{L_c(U)}
\end{equation}
according to Eq.~(\ref{U(L)}). In other words the manifold is rough up
to this length scale, but it is still smooth above:
\begin{equation}
  C(L) \approx 
\left\{ 
\begin{array}{ll}
\xi_0^2 [L/L_0]^{2 \zeta} & L < L_c,\\ 
\xi_0^2 [L_c/L_0]^{2 \zeta} & L > L_c.
\end{array}
\right.
\label{C(L)}
\end{equation}
Here the time dependence enters through $L_c \equiv L_c(U(t))$ via
Eqs.~(\ref{U(t).relax}) and (\ref{L_c(U)}). In the course of time the
manifold becomes rougher, the barriers increase and the dynamics
becomes even slower.

{\em Stationary driven state}. In contrast to the case without driving,
the manifold has bounded {\em effective} barriers or a finite {\em
  typical} barrier at finite velocity (maximum effective barrier). In
a {\em stationary} state, where the manifold is driven by a force $F$,
the typical barrier height $U$ determines the rate of jumps in
direction parallel or opposite to the driving force. The resulting
velocity $v$ is estimated by
\begin{equation}
  v \approx \frac{\xi}{t_0} \left[e^{- \frac{U - F
        \xi/2}{T}} - e^{- \frac{U +F \xi/2}{T}}
\right] \approx e^{-U/T} \mu F
\label{v(U,F)}
\end{equation}
for $F \xi \ll T$ with an effective mobility $\mu= \xi^2/ t_0 T$. The
scale $\xi$ is related to the size of the barrier and therefore
depends implicitly on $U$. However, this dependence is much weaker
than the exponential dependence displayed explicitly in
Eq.~(\ref{v(U,F)}) and can be ignored.

Inverting Eq.~(\ref{v(U,F)}) one can define from the stationary
velocity-force characteristics (VFC) $v_{\rm st}(F)$ a typical barrier
for stationary creep as a function of the driving force by
\begin{equation}
  U_{\rm st}(F)= T \ln [\mu F / v_{\rm st}(F)].
\label{def.U_st}
\end{equation}
The scaling theory predicts a dependence of the barrier height on the
driving force according to $U_{\rm st}(F) \sim F^{-\theta / (2 -
  \zeta)}$ \cite{scal}. 

The structure of the manifold in the driven state can be determined as
follows: the typical barriers are the maximum effective barriers,
which are able to pin local regions of the manifold {\em temporarily}.
The manifold will be rough according to Eq.~(\ref{C(L)}) only up to
the size of these regions. This size is related to the maximum
effective barrier $U_{\rm st}$ via Eq.~(\ref{L_c(U)}).

If the structure of the manifold initially does not correspond to the
stationary state, it will relax with a relaxation time $\tau(U_{\rm
  st})$ given by Eq.~(\ref{t(U)}), since $U_{\rm st}$ is the maximum
effective barrier. However, this time scale is relevant only very
close to the stationary state.

{\em Heuristic model}. We now develop a heuristic model describing
the manifold dynamics far away from the stationary state and the
response of the manifold to a time-dependent driving force.  In this
case the manifold needs a finite time to adapt its structure to the
varying force. For simplicity we describe the structure of the
manifold by a single parameter, the length scale $L_c(t)$ up to which
barriers are effective, or equivalently the corresponding energy scale
$U(t)$.  We assume that the velocity responds instantaneously to the
force according to
\begin{equation}
  v(t)= \mu F(t) e^{-U(t)/T}.
\label{v(t)}
\end{equation}
This means that the structure at the time $t$ determines only the
effective mobility at that moment.

Due to the finite relaxation time, which grows infinitely for the
decreasing velocity, the structure itself will {\em not} adapt itself
instantaneously to the driving force, but it will rather depend on the
history of the system.  We describe this history dependence effect by
a simple equation for the evolution of the typical activation barrier
$U(t)$:
\begin{equation}
  \frac{d}{dt} U(t)=\frac{T}{t_0} 
  \left( e^{-U(t)/T}-e^{-U_{\rm st}(F(t))/T} \right).
\label{eqmo}
\end{equation}
This is the central equation for our study of hysteretic creep. The
first term on the right hand side represents the tendency of the
manifold to get pinned by the disorder potential. Thereby it gains
potential energy and and the barriers increase. This tendency is
balanced by the second term, which incorporates a saturation of the
typical barrier due to thermally activated depinning of local regions
for finite force.

Although a microscopic derivation of Eq.~(\ref{eqmo}) is beyond our
capabilities, we would like to argue that it is the simplest
conceivable equation of motion of $U$ which satisfies the following
conditions: (i) in the absence of disorder one has an ohmic stationary
VFC with $U_{\rm st}(F)$. The equation of motion is then solved by
$U(t)=0$ independent of the driving force.  (ii) In the absence of a
driving force, $U(t)$ must grow logarithmically as in
Eq.~(\ref{U(t).relax}). This is indeed the case, since $e^{-U_{\rm
    st}(F=0)/T}$ vanishes in a glassy state. (iii) When $U(t)=U_{\rm
  st}(F(t))$, the manifold is in instantaneous equilibrium, where
$U(t)$ must not change. (iv) If the driving force $F>0$ is constant in
time, the structure has to approach its stationary state with a finite
relaxation time (\ref{t(U)}). This behavior is readily verified by
linearizing Eq.~(\ref{eqmo}) with respect to $U(t)-U_{\rm st}(F)$.
The Eq.~(\ref{eqmo}) may be also considered as a simplified version of
the evolution equation for a barrier distribution function in glassy
systems used in \cite{GIL93}.

{\em Phenomenology of the model}. In general, the hysteretic behavior
emerges as the system parameters (like the driving force and/or
temperature) change on a time scale of the order of or smaller than
the characteristic relaxation time.  Hence we expect these hysteretic
phenomena to be particularly pronounced in glassy systems, where the
equilibrium relaxation time diverges and aging effects occur.
However, since in the driven case the external force $F$ sets the
upper limit $U_{\rm st}(F)$ for the barriers the system can be trapped
into, the relaxation times are large and we expect noticeable effects
only at sufficiently small velocities.  As we shortly show the
equation of motion (\ref{eqmo}) is not only able to capture hysteretic
effects but even more it exhibits a retarded and almost threshold-like
onset of motion, when a driving force is switched on after the system
has been aged at zero driving force.

We performed a numerical integration of this differential equation for
some time-cycles of the driving force. These calculations have been
performed for a stationary characteristics with $U_{\rm st}(f)=T
f^{-0.5}$.  Here $f$ is a dimensionless force given by the ratio of
$F$ and the depinning force. The scaling exponent is that for the
three-dimensional vortex lattice with logarithmic roughness,
$\zeta=0$\cite{N90}. Time is measured in units of $t_0$.

\begin{figure}[t]
\epsfxsize=0.9 \linewidth
\epsfbox{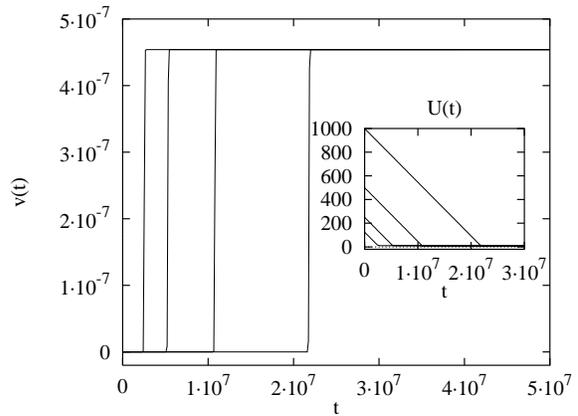}
\caption{
  Depinning of the manifold, when a constant driving force is switched
  on after aging at zero drive. Inset: Time evolution of the typical
  barrier $U(t)$. Main plot: Time evolution of velocity $v(t)$. The
  curves (from the left to the right) correspond to initial barriers
  $U(0)=$125, 250, 500, and 1000.}
\label{fig.depin}
\end{figure}

First we examine the time dependence of velocity in a depinning
scenario. It is supposed that the manifold has initially a large
typical barrier $U(t=0)$, which can be obtained by aging at zero
drive. Then at $t=0$ a finite drive $f>0$ is switched on such that
$U_{\rm st}(f) \ll U(0)$. This leads to a very abrupt onset of motion
illustrated in Fig.~\ref{fig.depin}.  This behavior can be understood
as follows: initially the barrier decreases linearly in time. This
implies that the velocity increases exponentially in time. The time
scale is set by the relaxation time of barriers of height $U_{\rm
  st}(F)$. After a time $t_{\rm depin} \approx \tau(U_{\rm st}(F))
U(t=0)/T$ the barrier height and drift velocity reach their stationary
values. On a linear velocity scale the exponential increase and the
saturation appear as an almost discontinuous onset of motion.

The sharpness of the onset of motion increases for smaller forces,
since there the stationary VFC has a stronger nonlinearity. In the
example of Fig.~\ref{fig.depin} the relaxation time $\tau(U_{\rm
  st}(f)) \approx 2 \cdot 10^5$ is smaller than $t_{\rm depin}$ by a
factor of order $10^2$. The depinning time $t_{\rm depin}$, which is
essentially proportional to the initial barrier, increases only
logarithmically as a function of the aging time, during which the
manifold has to be kept at zero drive in order to increase the initial
barrier height.

\begin{figure}[b]
\epsfxsize=0.9 \linewidth
\epsfbox{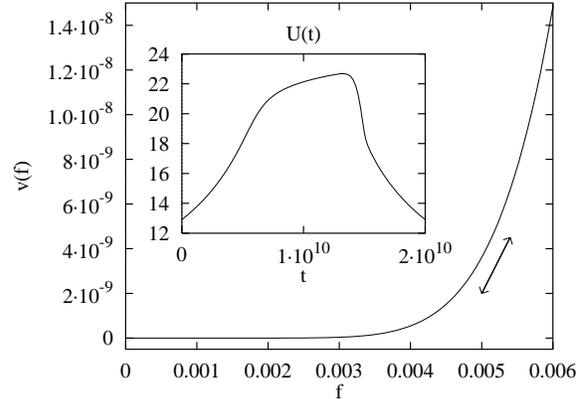}
\caption{
  The driving force is swept down from $f=6 \cdot 10^{-3}$ to $f=0$
  with constant rate in time $t=10^{10}$. Then it is immediately swept
  up again with the same rate. Initially $U=U_{\rm st}(f)$ is
  supposed. Inset: $U(t)$ during sweep down for $0<t<10^{10}$ and
  during subsequent sweep up for $10^{10} < t < 2\cdot 10^{10}$. Main
  plot: $v(f)$ shows no noticeable hysteresis.}
\label{fig.down}
\end{figure}

Next we perform sweeps of the driving force in time in order to
demonstrate the possibility of an extreme hysteresis in $v(f)$. As a
generic example we consider the case, where an initial force ($f_{\rm
  max}=6 \cdot 10^{-3}$) well below the depinning force ($f=1$) is
switched off with a constant rate (during a time $t_{\rm sweep}
=10^{10}$). The sweep time is chosen to satisfy $\tau(U_{\rm
  st}(f_{\rm max})) \ll t_{\rm sweep}$. The inset of
Fig.~\ref{fig.down} illustrates the time development of $U(t)$ during
this sweep. It increases first linearly and then logarithmically,
since finally the system ages as in the absence of a driving force.
Then the force is ramped up to $f_{\rm max}$ again with the same rate,
immediately after $f=0$ has been reached. The barrier continues to
grow until the driving force becomes so big that $U_{\rm st}(f(t)) >
U(t)$.  In a linear plot as in the main part of Fig.~\ref{fig.down},
the characteristics exhibits practically no hysteresis, i.e. $v(t)$
coincides with $v_{\rm st}(f(t))$.

\begin{figure}[t]
\epsfxsize=0.9 \linewidth
\epsfbox{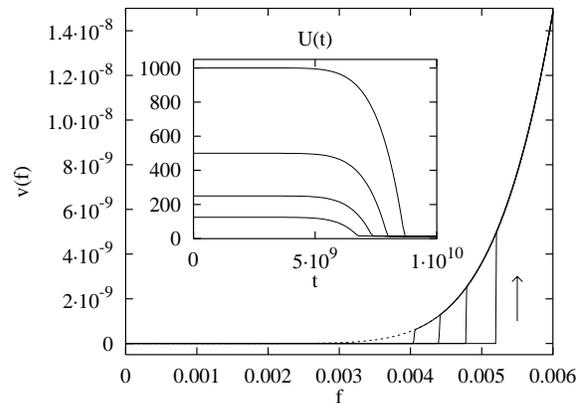}
\caption{
  The driving force is swept up from $f=0$ to $f=6 \cdot 10^{-3}$ with
  constant rate in a time $t=10^{10}$. Different initial conditions
  are used, see text. Inset: Time evolution of $U(t)$. The larger $U$
  initially, the later it comes down to the stationary value. Main
  plot: $v(t)$ versus $f(t)$. The larger $U$ initially, the later $v$
  jumps up to the stationary VFC (dashed line).}
\label{fig.up}
\end{figure}

Hysteresis appears only after the system has been aged at zero (or
very small) driving force. In the numerical example depicted in
Fig.~\ref{fig.up} we suppose that the system has been aged a certain
time until the barrier reaches values $U=125$, 250, 500, or 1000 (full
lines from the left to the right). These values correspond to
stationary barriers at forces of order $f \sim 10^{-6} \ll f_{\rm
  max}$. Starting from these barrier heights at $t=0$, the force is
switched on up to $f_{\rm max}$ linearly during $t_{\rm
  sweep}=10^{10}$. Initially the barriers decrease extremely slowly
and then faster and faster, until $U(t) \approx U_{\rm st}(f(t))$.
This again leads to a threshold-like onset of motion from an
effectively pinned state to a state, where the manifold moves with the
velocity $v(t)$ located on the stationary creep VFC.  As we increase
barriers $U(t)$ reflecting the growing waiting time at $f=0$, the
threshold force $f_{\rm th}(t)$ shifts upwards to larger values.  Note
the striking similarity of the obtained hysteretic threshold-like
onset of motion to the hysteretic switching that has been observed in
recent experiments on high-temperature superconductors\cite{exp}.

In conclusion, we have presented a simple phenomenological model of
elastic manifold driven through quenched disorder which displays
hysteretic dynamic behavior. Dynamic freezing of the system into deep
metastable states at zero driving force leads to a threshold-like
onset of motion at the threshold force $f(t_{\rm w})$ growing with the
waiting time $t_{\rm w}$ at $f=0$. It is important to note that the
presented threshold dynamics is {\it different} from the usual
depinning transition at {\it zero} temperature. The threshold-like
onset of motion demonstrated here is specific to finite temperatures
and occurs well below the depinning transition. It would be
interesting to perform a systematic study of the threshold dynamics in
the vicinity of the melting transition, where this dynamic hysteresis
interferes with the static hysteresis of melting proposed in
\cite{GIL93}.

We thank G. W. Crabtree, A. E. Koshelev, W. K. Kwok, and U.  Welp for
stimulating discussions. This work was supported from Argonne National
Laboratory through the U.S.  Department of Energy, BES-Material
Sciences, under contract No. W-31-109-ENG-38 and by the NSF-Office of
Science and Technology Centers under contract No.  DMR91-20000 Science
and Technology Center for Superconductivity.

\end{document}